\documentclass[format=acmsmall]{acmart}
\pdfoutput=1

\usepackage{booktabs} 

\usepackage[ruled]{algorithm2e} 

\setcopyright{rightsretained}

\usepackage[utf8]{inputenc}
\usepackage{url}
\usepackage{tikz}
\usepackage{xspace}
\usepackage{amsmath}
\usepackage{natbib}
\usepackage{mathtools}
\usepackage{multirow}
\usepackage{amsthm}
\usepackage{balance}
\usepackage{ctable} 

\makeatletter
\g@addto@macro\normalsize{%
  \setlength\abovedisplayskip{3pt}
  \setlength\belowdisplayskip{3pt}
  \setlength\abovedisplayshortskip{3pt}
  \setlength\belowdisplayshortskip{3pt}
}
\makeatother

\makeatletter
\newcommand*{\rom}[1]{\expandafter\@slowromancap\romannumeral #1@}
\makeatother

\usetikzlibrary{decorations.pathreplacing}
\usetikzlibrary{calc,shapes.multipart,chains,arrows,shapes.geometric,backgrounds,decorations.pathmorphing,positioning-plus}
\usetikzlibrary{arrows,arrows.meta}
\usetikzlibrary{positioning,shapes.misc} 
\pgfdeclarelayer{background}
\pgfdeclarelayer{foreground}
\pgfsetlayers{background,main,foreground}

\tikzset{
	headernode/.style = {
  	shape = rectangle,
    text depth = +0pt,
    draw,
    fill = white,
    rounded corners},
	basic box/.style = {
    shape = rectangle,
    align = center,
    draw  = #1,
    fill  = #1!25,
    rounded corners},
    split box/.style = {
    rectangle split,
    rectangle split horizontal=false,
    rectangle split parts=2,
	rectangle split part fill={#1!25,red!25},
    align = center,
    draw = #1},
  header/.style = {%
    inner ysep = +2em,
    append after command = {
      \pgfextra{\let\TikZlastnode\tikzlastnode}
      node [headernode, font = \large] (header-\TikZlastnode) at (\TikZlastnode.north) {#1}
      node [span = (\TikZlastnode)(header-\TikZlastnode)]
        at (fit bounding box) (h-\TikZlastnode) {}
    }
  },
  basic block/.style = {%
  	shape = circle,
    draw = black,
    fill = white,
    node distance = 0.5cm,
    minimum width = 0.5cm,
    font=\sffamily\small},
  bb record/.style={rectangle split ,rectangle split horizontal,
    rectangle split parts=#1,
    font=\small,
inner sep=4pt,
	fill = white,
    draw,text centered},
  hv/.style = {to path = {--++(0,-#1) -|(\tikztotarget)\tikztonodes}},
  hv simple/.style = {to path = {-|(\tikztotarget)\tikztonodes}},
  vh/.style = {to path = {|-(\tikztotarget)\tikztonodes}},
  fat blue line/.style = {ultra thick, blue},
  snake it/.style = {decoration={snake, 
    amplitude = .4mm,
    segment length = 2mm,
    post length=0.9mm},decorate}
}

\makeatletter
\tikzset{%
  prefix node name/.code={%
    \tikzset{%
      name/.code={\edef\tikz@fig@name{#1 ##1}}
    }%
  }%
}
\makeatother

 \newcommand{\AAABLOCK}{%
        \begin{tikzpicture}[inner sep=0pt,baseline=(base)]%
       	\node [bb record=1,inner sep=2pt]{\nodepart[text width=1.5cm]{one}A0-A1-A2};
        \node (base) at (0,-.5ex) {};
        \end{tikzpicture}%
    }

 \newcommand{\AABLOCK}{%
        \begin{tikzpicture}[inner sep=0pt,baseline=(base)]%
       	\node [bb record=1,inner sep=2pt]{\nodepart[text width=1.2cm]{one}A0$\,\to\,$A1};
        \node (base) at (0,-.5ex) {};
        \end{tikzpicture}%
    }
  
   \newcommand{\ABLOCK}{%
        \begin{tikzpicture}[inner sep=0pt,baseline=(base)]%
       	\node [bb record=1,inner sep=2pt]{\nodepart[text width=0.4cm]{one}A2};
        \node (base) at (0,-.5ex) {};
        \end{tikzpicture}%
    }
    
     \newcommand{\MAAABLOCK}{%
        \begin{tikzpicture}[inner sep=2pt,baseline=(base)]%
       	\node [bb record=2,inner sep=2pt]{\nodepart{one}M \nodepart[text width=1.5cm]{two}A0-A1-A2};
        \node (base) at (0,-.5ex) {};
        \end{tikzpicture}%
    }
    
    \newcommand{\MAAABBLOCK}{%
        \begin{tikzpicture}[inner sep=0pt,baseline=(base)]%
       	\node [bb record=3,inner sep=2pt]{\nodepart{one}M \nodepart[text width=1.5cm]{two}A0-A1-A2 \nodepart{three}B};
        \node (base) at (0,-.5ex) {};
        \end{tikzpicture}%
    }
    
     \newcommand{\CBLOCK}{%
        \begin{tikzpicture}[inner sep=0pt,baseline=(base)]%
       	\node [bb record=1,inner sep=2pt]{\nodepart{one}C};
        \node (base) at (0,-.5ex) {};
        \end{tikzpicture}%
    }

\newtheorem{layout}{Layout}%
\newtheoremstyle{plain}
  {\topsep}   
  {\topsep}   
  {\upshape}  
  {0pt}       
  {\bfseries} 
  {.}         
  {5pt plus 2pt minus 2pt} 
  {}          

\newtheorem{definition}{Definition}[section]

\newcommand\np{\mbox{\bf NP}\xspace}

\begin{document}

\title{Codestitcher: Inter-Procedural Basic Block Layout Optimization}

\author{Rahman Lavaee}
\authornote{This research was done when Rahman Lavaee was affiliated with University of Rochester}
\affiliation{
  \institution{Google}
  \city{Sunnyvale}
  \state{CA}
  \postcode{94089}
  \country{USA}
  }
\email{rahmanl@google.com}
\author{John Criswell}
\email{criswell@cs.rochester.edu}
\author{Chen Ding}
\email{cding@cs.rochester.edu}
\affiliation{%
  \institution{University of Rochester}
  \department{Computer Science}
  \city{Rochester}
  \state{NY}
  \postcode{14627}
  \country{USA}
}

\begin{abstract}
 Modern software executes a large amount of code.  Previous
  techniques of code layout optimization were developed one or two
  decades ago and have become inadequate to cope with the scale and
  complexity of new types of applications such as compilers, browsers, interpreters, language VMs and shared libraries.
  
  This paper presents Codestitcher, an inter-procedural basic block code layout optimizer which reorders basic blocks in an executable to benefit from better cache and TLB performance. Codestitcher provides a hierarchical framework which can be used to improve locality in various layers of the memory hierarchy. Our evaluation shows that Codestitcher improves the performance of the original program by 3\% to 25\% (on average, by 10\%) on 5 widely used applications with large code sizes: MySQL, Clang, Firefox, Apache, and Python. It gives an additional improvement of 4\% over LLVM's PGO and 3\% over PGO combined with the best function reordering technique.
\end{abstract}

\maketitle

\section{Introduction}
\label{section:intro}

For large applications, instruction misses are the main culprit for stalled cycles at the processor front-end. They happen not just in the instruction cache, but also in the unified cache at lower levels and in the TLB.\@ Increasing the cache and TLB capacities or their associativities results in lower miss rates, but also increases their access latencies. In particular, the on-chip L1 instruction cache and TLB are required to have very low hit latencies (1--4 cycles). As a result, they have experienced only slight increase in capacity and associativity. For example, from the Pentium \rom{2} to the Nehalem micro-architecture, the L1 instruction cache doubled in capacity (16KB to 32KB), but it retained its associativity of 4. From Nehalem to Haswell, it has retained its capacity, but doubled in associativity (4 to 8). Since Haswell, and until the most recent micro-architecture, Coffee Lake), the L1 instruction cache has not seen any improvement in capacity, associativity, or the block size~\cite{intel-manual}.

\setlength\tabcolsep{4pt} 
\begin{table}[h!]
\caption{Code size growth for MySQL and Firefox, in terms of LOC (lines of C/C++ code), and \texttt{text} size.}
\label{tab:code-size}
\begin{tabular}{|c|c|c|c|c|}
\hline
application & version & release year & LOC & \texttt{text} size \\ \hline
MySQL & \begin{tabular}{c} \multirow{2}{*}{} 6.2.16 \\ 7.5.8 \end{tabular} &
\begin{tabular}{c} \multirow{2}{*}{} 2008 \\ 2017 \end{tabular} &
\begin{tabular}{c} \multirow{2}{*}{} 0.9M \\ 2.1M \end{tabular} &
\begin{tabular}{c} \multirow{2}{*}{} 3.6MB \\ 13.6MB \end{tabular} \\
\hline
Firefox & \begin{tabular}{c} \multirow{2}{*}{} 1.0.1 \\ 52.0 \end{tabular} &
\begin{tabular}{c} \multirow{2}{*}{} 2004 \\ 2017 \end{tabular} &
\begin{tabular}{c} \multirow{2}{*}{} 2.4M \\ 7M \end{tabular} &
\begin{tabular}{c} \multirow{2}{*}{} 7.2MB \\ 41.7MB \end{tabular} \\
\hline
\end{tabular}
\setlength\tabcolsep{6pt} 
\end{table}

In the meantime, modern software is growing in code size and complexity. Table~\ref{tab:code-size} shows the increase in code size in terms of lines of code (LOC) and the program's binary size, for two applications: MySQL and Firefox. Assuming a fixed yearly rate in code growth, MySQL has grown by 10\% in code size each year, while Firefox has grown by 9\%. However, as language constructs become more complex each year, the yearly increase in LOC translates to a larger rate of increase in text size: MySQL and Firefox have grown respectively by 16\% and 14\% in text size, every year.

For code, longer cache lines seem more profitable as code is immune to false sharing and is more likely to be sequential. However, using a longer cache line only for code is not practical because of the inclusive, unified caches at lower levels. In modern architectures, hardware prefetching and stream buffers exploit the spatial locality beyond a single cache line.

Considering the slow rate of improvement in capacities of on-chip caches, it is natural to wonder how we can utilize them in an optimal way. An important factor is code layout. It affects the instruction cache performance in several ways. First, cache lines that are shared between unrelated code segments (functions or basic blocks which do not execute together) may result in lower utilization of the cache. Second, when these code segments are stored consecutively in memory, prefetching may fill the cache with useless code. Third, in a set-associative cache, code blocks that are mapped to the same set may result in conflict misses. Finally, unrelated code segments which reside in the same page may inflict pressure on instruction TLB.\@

Profile-guided code layout optimization has been carefully studied in the past. 
An influential solution was proposed by Pettis and Hansen~\cite{PettisH:PLDI90} (called PH in short).\@ PH consists of function splitting, intra-function basic block chaining, and global function ordering.
For Function splitting, PH separates the hot code from the cold code by splitting every function into these two parts. Next it performs basic block chaining to coalesce the hot part of every function. These function parts are then passed to the function reordering stage where PH finds an optimal ordering based on a greedy heuristic.

For over two decades, code layout optimization techniques have mainly followed the three-step framework suggested by PH.\@ The research has primarily been focused on suggesting new heuristics for the function reordering stage. The main two shortcomings of prior techniques are as follows.

\begin{enumerate}
\item The coarse-grained hot-cold function splitting limits the scope of the layout optimization.
\item Function reordering heuristics are not based on precise spatial distance between related code segments.
\end{enumerate}

To overcome these shortcomings, we introduce \mbox{{\bf Codestitcher}}, a new framework for inter-procedural basic block reordering. Unlike prior techniques, Codestitcher splits functions into as many parts as necessary to expose all opportunities for placing related code segments together in memory. More importantly, Codestitcher uses the layout distance between related instructions as a parameter to incrementally construct the overall layout.  Using this parameter, code collocation proceeds hierarchically, maximizing its benefits within successive layers of the memory hierarchy.

The main contributions of this paper are as follows:

\begin{itemize}
\item We identify the locality improvement opportunities that an inter-procedural basic block layout can offer, compared to a per-function layout.

\item We present a new solution for basic block chaining, based on maximum cycle cover. Unlike prior techniques, our solution offers a theoretical guarantee.

\item We formally define the distance-sensitive code layout problem. Then, we 
present a hierarchical framework which allows us to optimize code locality in all layers of the memory hierarchy, by successively solving the distance-sensitive code layout problem for various distance parameters. 

\item We present a careful investigation into how we can choose the distance levels for a typical hardware platform. We also give insights into the implications of using large pages for code.

\item We show how branch prediction is affected by inter-procedural basic block layout and why branch directions should not hamper locality optimizations.

\item We present two implementations of Codestitcher; an instrumentation-based version and a version based on Linux perf utility.

\item Finally, we present an evaluation of both implementations of Codestitcher on five widely-used programs with large code sizes, and on two hardware platforms. We also analyze separately, the effect of using large pages, and code layout granularity (function vs. basic block).

\end{itemize}

The rest of the paper is organized as follows.  Section~\ref{section:design} describes the design of Codestitcher and motivates its design.  Section~\ref{section:impl} discusses our prototype implementation of Codestitcher in LLVM.\@  Section~\ref{section:results} presents our evaluation on a set of five widely-used programs, besides explaining the implementation of the perf-based Codestitcher. Section~\ref{sec:related} discusses related work,  and Section~\ref{sec:summary} concludes.

%


\section{Design}
\label{section:design}

\subsection{Overview}
\label{sec:overview}

\begin{figure}[h]
\includegraphics[scale=0.6]{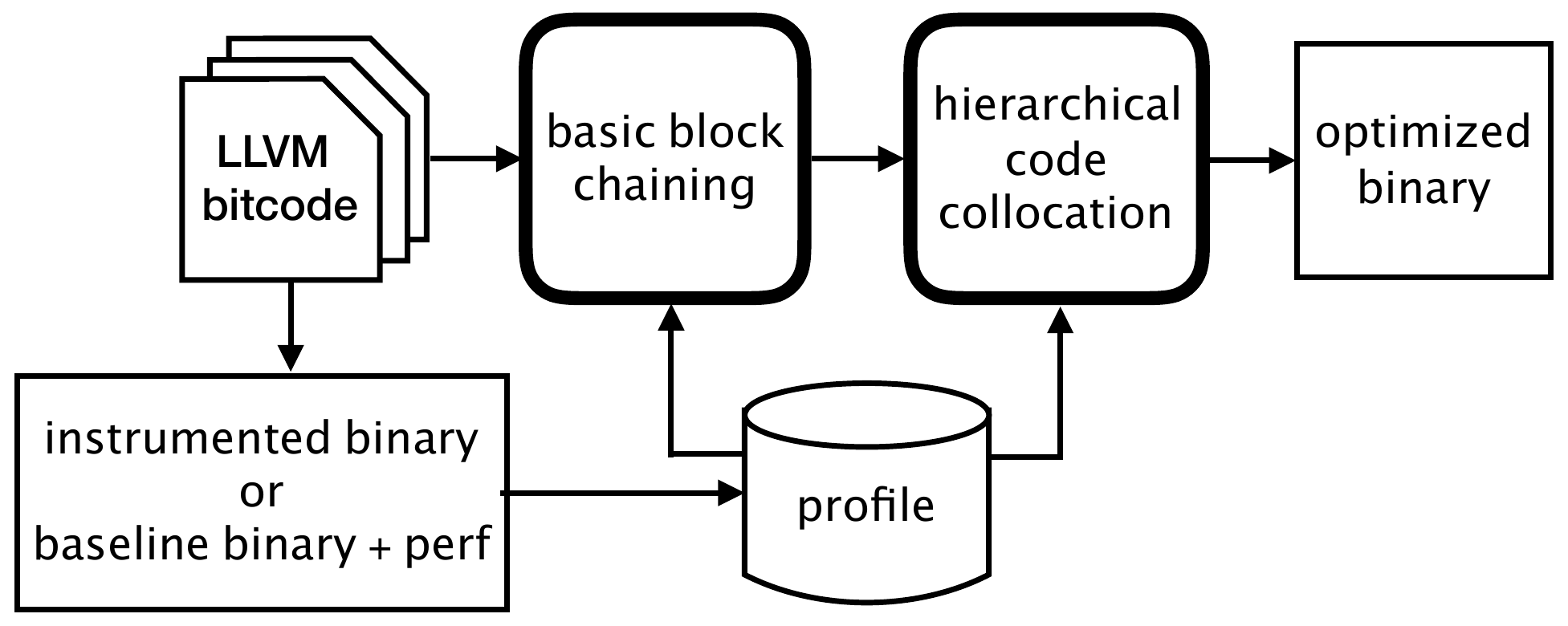}
\caption{High-level overview of Codestitcher}
\label{fig:overview}
\end{figure}

Figure~\ref{fig:overview} gives a high-level overview of Codestitcher. The source code is first compiled to obtain LLVM bitcode files and the instrumented binaries. Profiles are then collected by running the instrumented program. Alternatively, Codestitcher can skip instrumentation and use the Linux perf utility to sample profiles from  the execution of a symbolized version of the baseline program.

After collecting the profiles, Codestitcher performs \emph{Basic Block Chaining} based on the profile data, to minimize the number of unconditional jumps that the program executes. The result of this step is a sequence of inter-procedural basic block (BB) chains. A BB chain is a sequence of basic blocks which terminates with an unconditional branch or return, but such instructions cannot happen in the middle of a BB chain. In this paper, functions calls are not considered to be unconditional jumps.

The constructed basic block chains are then passed to the \emph{Hierarchical Code Collocation} stage, where Codestitcher iteratively joins them together to form longer sequences of basic blocks (basic block layouts). Conceptually, in this stage, Codestitcher iterates through a number of code distance parameters. For each distance parameter, Codestitcher collocates code segments in order to maximize spatial locality within that distance.

\subsection{Motivations}
\label{sec:mot}

\subsubsection{Inter-Procedural Basic Block Layout}
\label{sec:mot:inter}
\begin{figure}[h]
  \centering
  \begin{tikzpicture}[->, >=stealth, font=\sffamily\small , node distance=1cm]
  	\tikzstyle{func} = [draw,fill=none,circle]

  \node (M) [func] {M};
  \node (A0) [right =0.8cm of M] {A0};
  \node (A1) [above right =0.2cm and 0.6cm of A0] {A1};
  \node (A2) [below right =0.2cm and 0.6cm of A0] {A2};
  \node (B) [func,right = 0.8cm of A1] {B};
  \node (C) [func,right = 0.8cm of A2] {C};
  \node () at (1.4,1) {A};
  
  \path (M) edge node [above] {100} (A0)
  		(A0) edge [densely dashed] node [above =0.06 cm] {80} (A1)
        	edge [densely dashed] node [below =0.06 cm] {20} (A2)
        (A1) edge node [above] {80} (B)
        (A2) edge node [below] {20} (C);
    
  \draw (2.2,0) circle (1.1cm);
\end{tikzpicture}
\caption{Weighted inter-procedural control flow graph of a program, over a hypothetical execution. Every circle represents a function. Control flow edges are represented by dashed lines and call edges are represented by solid lines.}
\label{fig:cg}
\end{figure}
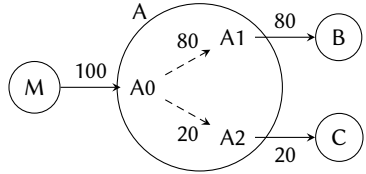

We use a contrived example to show intuitively why an inter-procedural basic block layout can deliver higher improvements compared to a per-function layout. Consider the inter-procedural control flow graph in Figure~\ref{fig:cg} where function M calls function A 100 times during the program execution.
The entry basic block of A (A0) ends with a conditional branch. The branch jumps to A1 80 times ands jumps to A2 20 times. A1 calls function B but A2 calls function C.

A per-function layout for this problem requires merging all basic blocks of A before function reordering. PH, for example, performs BB chaining to form the layout \AAABLOCK\, for A. Next, PH processes the call frequency edges in decreasing weight order. When processing each edge, PH joins together the layouts connected by that edge. A possible final layout is shown below.

\begin{layout}
\label{layout:ph}
\begin{tikzpicture}[inner sep=0pt,baseline=(base)]
\node (G3-M-A-B-C) [bb record = 4, anchor=west] {\nodepart{one}M \nodepart[text width=1.5cm]{two}A0-A1-A2 \nodepart{three}B \nodepart{four}C };
\node (base) at (0,-.5ex) {};
\end{tikzpicture}
\end{layout}


The quality of a code layout can be evaluated with locality measures. A simple measure for code locality within one function is sequential code execution. Similarly, as a simple inter-procedural metric, we can look at how often control transfers between adjacent basic blocks.

In our example program, a total of 300 control transfers happen between 6 basic blocks and via 5 call and control flow edges.
In layout~\ref{layout:ph}, only two edges (M$\,\to\,$A0 and A0$\,\to\,$A1) run between neighboring blocks (a total of 180 control transfers). Evidently, forming a single code segment for A has led both calls in A to separate from their callees (B and C). 

On the other hand, an inter-procedural basic block layout can benefit from a finer grain splitting for $A$. In particular, if $A$ is split into hot code paths ( \AABLOCK\, and \ABLOCK\, ), all function entries can be attached to their caller blocks, as shown in the layout below.
\begin{layout}
\label{layout:stitch}
\begin{tikzpicture}[inner sep=0pt,baseline=(base)]
\node (G3-M-A-B-C) [bb record = 5, anchor=west] {\nodepart{one}M \nodepart[text width=1.2cm]{two}A0$\,\to\,$A1 \nodepart{three}B \nodepart{four}A2 \nodepart{five}C };
\node (base) at (0,-.5ex) {};
\end{tikzpicture}
\end{layout}



With Layout~\ref{layout:stitch}, all edges except A0$\,\to\,$A2 run between adjacent blocks. This gives a total of 280 control transfer between adjacent blocks, which is an increase of about 55\% over PH.\@

\subsubsection{Layout Distance Between Related Instructions}

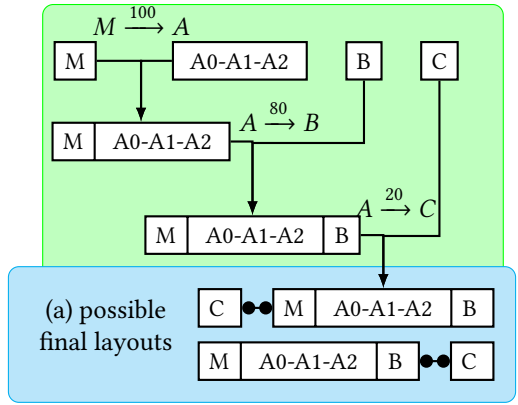
\begin{figure}[h]
\centering
\begin{tikzpicture}[thick, nodes = {align = center},
    >=latex]

 \node (phantom) {};
       \node (G3-M) [bb record=1, below  = 0.1cm of phantom] {M};
  \node (G3-A) [bb record= 1, right =1cm of G3-M] {\nodepart[text width=1.5cm]{one}A0-A1-A2};

    \node (G3-B) [bb record=1 ,right = 0.5cm of G3-A] {B};
   \node (G3-C) [bb record=1,right = 0.5cm of G3-B] {C};

    \node (G3-M-A) [bb record = 2, below = 0.8cm of ($(G3-M)!0.4!(G3-A)$)] {\nodepart{one}M \nodepart[text width=1.5cm]{two}A0-A1-A2};
  
   
      \node (G3-M-A-B) [bb record = 3, below = 1.5cm of ($(G3-M-A)!0.5!(G3-B)$)] {\nodepart{one}M \nodepart[text width=1.5cm]{two}A0-A1-A2 \nodepart{three}B};

    \node (G3-M-A-B-1) [bb record = 3, below = 2.3cm of ($(G3-M-A-B)!0.7!(G3-C)$)] {\nodepart{one}M \nodepart[text width=1.5cm]{two}A0-A1-A2 \nodepart{three}B };

        \node (G3-C-1) [bb record = 1, left=0.4cm of G3-M-A-B-1] {\nodepart{one}C };

     \node (G3-M-A-B-2) [bb record = 3, below = 0.7cm of G3-C-1.west, anchor=west] {\nodepart{one}M \nodepart[text width=1.5cm]{two}A0-A1-A2 \nodepart{three}B };
     
     \node (G3-C-2) [bb record = 1, right=0.4cm of G3-M-A-B-2] {\nodepart{one}C };

    \path[hv=0] (G3-M.east) edge [->]  (G3-M-A); 
      \path[hv=0] (G3-A.west) edge [->] node [above = 2mm ] {$M\xrightarrow{100} A$} (G3-M-A); 
            \path[hv=0] (G3-M-A.east) edge [->] node [above right= 0mm and -3mm] {$A\xrightarrow{80} B$} (G3-M-A-B); 
            \path[hv=1.08] (G3-B) edge [->] (G3-M-A-B); 
            \path[hv=0] (G3-M-A-B.east) edge [->] node [above right = 1mm and -5mm] {$A\xrightarrow{20} C$} (G3-M-A-B-1); 
            \path[hv=2.042] (G3-C.south) edge [->] (G3-M-A-B-1); 
  
  		  \draw[{Circle}-{Circle}] (G3-C-1) -- (G3-M-A-B-1);
          \draw[{Circle}-{Circle}] (G3-M-A-B-2) -- (G3-C-2);
   
         \begin{scope}[on background layer]
       \node[fit = (phantom)(G3-M)(G3-A)(G3-B)(G3-C)(G3-M-A)(G3-M-A-B)(G3-M-A-B-1)(G3-C-1)(G3-M-A-B-2)(G3-C-2) , basic box = green] (collocation) {};
     \end{scope}
     
       \node (collocation-result-label) [left = 0.5cm of G3-C-1.south] {(a) possible\\final layouts};
     \begin{scope}[on background layer]
        	\node[fit=(collocation-result-label)(G3-M-A-B-1)(G3-C-1)(G3-M-A-B-2)(G3-C-2), basic box=cyan, inner sep=8pt] (collocation-result) {};
            \end{scope}
\end{tikzpicture}
\caption{Pettis-Hansen function reordering, applied on the program in Figure~\ref{fig:cg}}
\label{fig:ph-reordering}
\end{figure}

Common procedure reordering heuristics follow a bottom-up approach. Starting with an initial set of code segments (hot function parts in PH), they iteratively join code segments together to form longer code segments. At each step, the heuristic makes two choices: which two code segments to join and in which direction. For instance, PH joins the code segments which are most heavily related by call frequency edges. In our example program in Figure~\ref{fig:cg}, PH first joins M with A to form the layout \MAAABLOCK. This layout then joins with B to form \MAAABBLOCK. Finally, connecting this layout with C gives the optimal PH layout. The three steps are shown in Figure~\ref{fig:ph-reordering}.

At the last step, when \MAAABBLOCK\, and \CBLOCK\, join together, PH faces a choice between different merging directions (as shown in Figure~\ref{fig:ph-reordering}(a)). The PH heuristic focuses on the merging points and chooses the direction that results in the largest number of calls between the merging points. In this example, no calls happen between the merging points (C vs. M and B). Therefore, PH treats both directions equally beneficial. 
However, we notice that the lower orientation (also shown in Layout~\ref{layout:ph}) forms a closer distance between A2 and C, which means a larger improvement in code locality.

Strikingly, the layout distance between related instructions (instructions connected by control flow edges) can guide us in making the best choice at every iteration. With each (ordered) pair of code segments, and every code distance level, we can attribute a value which indicates the number of control transfers which would happen within that distance if those two code segments join with each other. Then rather than solely focusing on calling frequencies, we can maximize the total number of control transfers which happen within close distances.

\subsubsection{Basic Block Chaining via Maximum Path Cover}
\label{sec:fallt}

Reordering the program's basic blocks
may require inserting jump instructions to conserve the control flow. Clearly, this is not the case for a function layout. This means function reordering can easily be implemented in a linker. For instance, the Darwin linker supports function ordering via the command line flag \texttt{-order\_file}. However, as we argued in Section~\ref{sec:mot:inter}, inter-procedural basic block layouts disclose new opportunities for improving code locality. The challenge is how to split functions
without adding extra jumps. 

The execution cost of unconditional jumps in modern processors is minimal thanks to the deep execution pipeline. However, additional jumps increase the instruction working set, which in turn increases the pressure on instruction cache and other layers of the memory hierarchy. As a result, minimizing the number of executed unconditional jumps is the first step towards finding the optimal layout. 

Minimizing the number of executed unconditional jump instructions is equivalent to maximizing the number of fall-throughs. That is, the total number of times execution falls through to the next basic block. We formalize the fall-through maximization problem as follows.
\begin{definition}[Fall-Through Maximization Problem]
	Given the control flow graph $G$ for the whole program, along with frequencies of each control flow edge in $G$, find a decomposition of $G$ into a set of disjoint chains $L$ which maximizes the total frequencies of edges in $L$.
\end{definition}
This problem is equivalent to the problem of \emph{maximum weighted path cover} in control flow graphs. The general maximum weighted path cover problem is \mbox{\np-hard}. The simple greedy approach used by PH is the most well-known heuristic for solving this problem. This solution does not give any theoretical guarantee, but has a quick running time of $O(m\lg n)$ on a graph with $m$ edges and $n$ vertices. On the other hand, a direct $1/2$-factor approximation algorithm exists for this problem. It is described as follows.


Given the weighted directed graph $G(V,E)$, we first remove all self-loops from $G$. Then we add zero weight edges for all non-existing edges in the graph (including self-edges) and find a \emph{maximum cycle cover} in the resulting graph. A maximum cycle cover is a set of disjoint cycles which covers all vertices and has the maximum weight among all cycle covers. The maximum cycle cover can be reduced to the problem of maximum  matching in weighted bipartite graphs. Thus, for a function with $n$ hot basic blocks and $m$ hot edges, we can use the Hungarian algorithm~\cite{DKuhn10} to find the maximum cycle cover in time $O(n^2m)$. 

After finding the maximum cycle cover, we convert it to a path cover by dropping the lightest edge from each cycle. It is easy to verify that the weight of the resulting path cover is at least within a factor 1/2 of the optimal path cover. 

We further improve the approximate path cover solution using the same idea as the greedy approach. That is, after finding the approximate path cover, we consider the control flow edges in decreasing weight order, and add them to the path cover if they connect the tail of one path to head of another.

The approximate solution has a theoretical bound but is not always guaranteed to deliver a better path cover than the greedy solution. In our experiments, we find that although the total number of fall-throughs for the whole program is higher for the approximate solution, the greedy solution outperforms the approximate for some functions. Therefore, we combine the two solutions to generate one that beats each individual solution.

\subsubsection{Tail Call Fall-Throughs}

Fall-throughs are traditionally defined as the transfer of control from one basic block to its next within the scope of a single function. However, this definition ignores fall-throughs which happens across functions. 
For regular calls, fall-through is not beneficial as it introduces additional jumps at the callsite. However, this is not the case for tail calls.

A tail call is a special type of call where the call is the last operation of the calling function.  A compiler can transform the code, so the callee does not return to the caller after the tail call, and the tail call is changed to a jump instruction.  Therefore, we can treat \emph{tail calls} and \emph{tail jumps} as interchangeable terms.

If function F calls function G via a tail call, we can remove the call instruction at F by placing the entry basic block of G right after the the tail call at F. Thus basic blocks which end with tail calls can form inter-procedural BB chains by chaining with the entry basic block of their callee functions. However, as we mentioned above, it is not beneficial to join basic blocks, inter-procedurally at regular callsites.
Tail call may also run within the same function, and are called recursive tail calls. These calls can be removed in a similar fashion. 

It is crucial to note that removing tail calls is only possible in an inter-procedural basic block layout. In a per-function code layout, it is not possible to remove tail jumps because they may reside in the middle of a function and removing them may alter the program's CFG.\@

\subsection{Hierarchical Merging of Code Segments}
\label{sec:hier}
Spatial locality is a multi-dimensional concept~\cite{Gupta:IPDS13}, defined as the \emph{frequent} and \emph{contemporaneous} use of data elements stored in \emph{close storage locations}. Two dimensions are time and space. That is, how close in time the data elements are accessed, and how close in memory the elements are stored. A third dimension is frequency: how frequent those accesses are.
The challenge in data/code layout optimization is to respect the importance of all three dimensions at the same time.

For code layout, unraveling the time and frequency dimensions is more tractable, as the program execution precisely overlaps with instruction accesses. Instructions within a basic block respect the spatial locality. Therefore, code layout optimizations only focus on control transfers across basic blocks (branches, calls, and returns).

The real challenge is the space dimension because it is affected by code layout. The challenge becomes more significant in a finer grain code layout. For example, the distance between branch instructions and their targets is fixed in a function layout, but can vary in a basic block layout. The goal of code layout optimization is to ensure that frequently executed control instructions travel a smaller distance in the program binary. 

Let us formally define the concept of layout distance. In our discussions below $i$ and $j$ always refer to basic blocks of the program.

\begin{definition}[Layout Distance]
\label{def:dist}
We denote by $dist(i,j)$ the number of bytes from $i$ to $j$ in the code layout, including $i$ and $j$ themselves.
\end{definition}

For a distance parameter $d$, and a control transfer $i\rightarrow j$, we call $i\rightarrow j$ a \emph{$d$-close transfer} if $dist(i,j) \leq d$. We also denote by $f(i,j)$ the number of times control transfers from $i$ to $j$. If control never transfers from $i$ to $j$, $f(i,j)=0$.

We now define the \emph{$d$-close Code Packing} problem.

\begin{definition}[$d$-close Code Packing]
\label{def:pack-full}
Order the program's basic blocks in a way that maximizes the total number of $d$-close transfers. That is, 
\[ \sum_{\substack{i,j \\ dist(i,j)\leq d}} f(i,j).\]
\end{definition}


Solving this problem for a specific distance $d$ would result in optimal spatial locality, but only in a limited distance. Given the hierarchical design of memory systems, it is important to improve spatial locality within different memory layers.


Codestitcher can follow this hierarchical design by successively solving the $d$-close code packing problem for increasing distance levels. At each distance level, Codestitcher gets as input an initial \textbf{partial layout}. A partial layout is a set of BB sequences, where each basic block appears in exactly one sequence. Codestitcher then solves the code packing problem for this distance level and passes the new partial layout to the next level. 

The exact formulation of the problem Codestitcher solves at every distance level is a bit different from Definition~\ref{def:pack-full}. Let us introduce some terms that will help us in describing this problem and its solution. 

\paragraph{\bf Partial Layout Distance}  In a partial layout $L$, 
for any two basic blocks $i$ and $j$, their partial layout distance, $dist_L(i,j)$, is the same as in Definition~\ref{def:dist}, except that if $i$ and $j$ belong to different BB sequences, then their distance is $\infty$.

\paragraph{Super-Layouts and Sub-Layouts}
Let $L_1$ and $L_2$ be two partial layouts. $L_2$ is a super-layout for $L_1$ if it can be derived from $L_1$ by joining some (or none) of the BB sequences in $L_1$. In that case, $L_1$ is called a sub-layout of $L_2$. A proper sub-layout of $L$ is  said to be of \emph{finer granularity} than $L$.

\paragraph{$d$-close transfers in a partial layout}
Let $L$ be a partial layout. We define $t_d(L)$, the total number of $d$-close transfers in $L$, as 
\[ t_d(L) = \sum_{S \in L} \sum_{\substack{i,j \in S \\ dist_{L}(i,j)\leq d}} f(i,j).\]
In other words, $t_d(L)$ is the sum of $d$-close control transfers over all BB sequences in $L$.

We now define the problem of $d$-close partial layout, the building block of Codestitcher.

\begin{definition}[$d$-close partial layout]
Let $L_0$ be the initial partial layout.
Find the \emph{finest grain} super-layout of $L_0$ which has the \emph{maximum} number of $d$-close control transfers ($t_d(L)$).
\end{definition}

The finer granularity constraint prevents joining BB sequences unless doing so results in additional $d$-close transfers. Effectively, it helps the next distance levels benefit from a larger number of close transfers.


\subsection{Solving the \textit{d}-Close Partial Layout Problem}
\label{sec:solve}

The $d$-close partial layout problem asks for an optimal super-layout for $L_0$, that is, a super-layout with the maximum number of $d$-close transfers. 
Let $L$ be a typical super-layout for $L_0$. Each BB sequence $S \in L$ is the result of joining some BB sequences in $L_0$. Let $I(S)$ be the set of those sequences.

The initial partial layout is fixed. Therefore, maximizing $t_d(L)$ is equivalent to maximizing its additive value with respect to $L_0$, which is $\Delta t_d(L) = t_d(L)-t_d(L_0)$. This value can be expressed as follows.
\begin{equation}
\label{eq:obj}
\Delta t_d(L) = \sum_{S \in L}\quad \sum_{\substack{Q,R \in I(S)\\ Q\neq R}}\, \sum_{\substack{i \in Q , j \in R \\ dist_{L}(i,j) \leq d}} f(i,j). 
\end{equation}
$\Delta t_d(L)$ is the objective value of the $d$-close partial layout problem. Its expression is intimidating, but easy to explain. The summation goes over all pairs of basic blocks $i$ and $j$ which didn't belong to the same BB sequence in the initial layout ($L_0$), but do in $L$.

We solve the $d$-close partial layout problem as follows.

For each BB sequence $S$ and every basic block $i\in S$, we define $F(i)$, the forward position of $i$ as the number of bytes from the beginning of $S$ to right after $i$. We also define $B(i)$, the backward position of $i$, as the number of bytes from the end of $S$ to right before $i$. 

We now define a directed weighted graph $G$ on the set of BB sequences in $L_0$. For each two BB sequences $S,T \in L_0$, we set the weight of the $(S,T)$ edge equal to the number of $d$-close transfers between $S$ and $T$ when $T$ is positioned right after $S$ in the final layout. More formally, we have

\begin{equation} \label{eq:gp}
w(S,T) = \sum_{\substack{i \in S , j \in T \\ B(i)+F(j) \leq d}} {f(i,j) + f(j,i)}.
\end{equation}

A super-layout for $L$ is equivalent to a path cover for $G$. 
However, unlike the fall-through maximization problem, here, non-adjacent BB sequences may contribute to value of $t_d(L)$. (In equation~\ref{eq:obj}, $Q$ and $R$ may be any two BB sequences in $I(S)$, not just adjacent ones.) Therefore, the weight of this path cover is only a lower bound on $\Delta t_d(L)$. Nevertheless, we can still use the graph formulation to compute a greedy solution.\@ We describe the algorithm as follows.

At every step, we find the heaviest edge in $G$. Let it be $(S,T)$. We connect the instruction sequences corresponding to $S$ and $T$, and replace $S$ and $T$ with a new node representing the joined sequence $S.T$. Then we insert edges connecting this new node to the rest of the graph according to Formula~\ref{eq:gp}. We continue this process until the additive value cannot be further improved. In other words, all edges in $G$ become of zero weight. 

Applying the greedy approach as explained above has a disadvantage. The heaviest edges in $G$ are more likely to run between longer instruction sequences. Joining these long instruction sequences together may prevent higher gains in subsequent iterations and distance levels. To solve this problem, we set the weight of each edge $(S,T)$ equal to  $\frac{w(S,T)} {size(S)+size(T)}$, where $size(S)$ and $size(T)$ are respectively the binary sizes of $S$ and $T$.\@ The new edge weight formulation allows us to join shorter instruction sequences before longer ones. 


We implement this algorithm as follows. First, we compute the inter-procedural control flow graph. That is, for each control flow instruction, a list of instructions it can jump to, along with the frequency of each control transfer. We use the control flow graph to build the directed weighted graph $G$ as we explained above. Then we build a max heap of all edges $\langle S,T \rangle \in G$ which have nonzero weights. This max heap helps us efficiently retrieve the edge with highest weight density at every iteration. At every iteration of the algorithm, we use the control flow graph along with the current location of instructions to update $G$, while we also update the max heap accordingly.

\subsection{Choosing the Distance Levels}
The initial partial layout for the collocation stage is the output from basic block chaining. Effectively, basic block chaining can be viewed as the 0-close partial layout problem. The precise equivalence requires an alternative definition of  layout distance. However, intuitively, BB chaining improves the spatial locality within the zero distance (fall-throughs).

To maximize the locality improvement, the distance levels must be chosen carefully. Next, we argue how we can obtain this list for our hardware platform, with special care towards two of the most important components of the architecture: cache and TLB.\@


The CPU fetches instructions by accessing the L1 instruction cache and TLB.\@ The L1 instruction cache is indexed by virtual addresses. This allows translation and cache lookup to be performed in parallel. The L1 instruction cache uses a standard 32~KB, 8-way set associative configuration with 64 byte lines. This means that the 64 memory blocks in each (4~KB) page are mapped one-to-one to 64 congruence classes in the cache. Therefore, we set the first distance level equal to 4096, the size of a regular page. Increasing close transfers within this distance level has two benefits.

\begin{itemize}
\item First, it minimizes the transfer of control between different pages, thereby, improving the performance of the instruction TLB.\@
\item Second, it reduces the conflict misses in instruction cache.  Specifically, if two instructions are less than 4~KB apart, they are guaranteed to be mapped to different sets.
\end{itemize}

The L2 and L3 caches are respectively 256~KB and 3~MB, with associativities of 8 and 12. Therefore, they are respectively 32~KB and 256~KB ``tall''\footnote{It is the aggregate size of a single cache way.} (512 and 4096 congruence classes). One may wish to apply Codestitcher at these distance levels. However, since these caches are physically indexed, the virtual to physical address mapping implies that a virtual page could be mapped to any of the 8 (or 64) page-aligned set of congruence classes in the L2 (or L3) cache. As a result, with 4~KB pages, such distance upper bounds (32~KB and 256~KB) cannot guarantee that memory blocks will be mapped to different cache sets. However, with 2MB pages, it is guaranteed that instructions in a 32~KB (or 256~KB) chunk of code are not mapped to the same L2 cache set. Thus applying Codestitcher on these distance levels can help if 2MB pages are used. 

In a similar fashion, we can capture the distance levels appropriate for improving the TLB performance. The TLB hierarchy in the Haswell architecture consists of a dedicated L1 TLB for instruction and data caches, plus a unified TLB for L2. The L1 instruction TLB has 128 entries for 4~KB pages, with an associativity of 4, along with 8 fully associative entries for large pages. The L2 TLB (STLB) has 1024 entries shared by 4~KB and 2~MB pages, and an associativity of 8. TLBs are always virtually indexed. Thus for TLBs we don't face the problem we discussed above. Focusing on small pages, a distance upper bound of 128~KB guarantees that page translation entries are mapped to different L1 TLB sets. Analogously, an upper bound of 512~KB on instruction distance guarantees that the page translation entries for those instructions are mapped to different sets of STLB.\@

Overall, the distance levels are as follows.
\begin{itemize}
\item When using regular 4~KB pages: 4~KB, 128~KB, 512~KB.
\item When using large 2~M pages: 4~KB, 32~KB, 256~KB, 2~MB.
\end{itemize}

In this paper, we perform separate evaluations when using regular and large pages for code. However, rather than generating separate layouts for each of the distance level lists above, we construct a single layout by combining the two.
The combined distance level list is 4~KB, 32~KB, 128~KB, 256~KB, 512~KB, 2~MB.\@
With this list, a distance parameters may not be beneficial for one page size, but can give a sub-optimal solution for the next distance level (the next memory layer). For instance, when using regular pages, code collocation within a 32KB distance does not result in improvements in L2 cache, but can improve the instruction TLB.\@

\subsection{Interaction With Branch Prediction}
\label{sec:br}
Traditionally, the performance of branch prediction has had a less significant impact on the design of code layout techniques, compared to code locality.
Moreover, the use of more informative branch history tables has helped modern processors improve their branch prediction performance, to the extent that code layout can only impact the static branch prediction. That is, when branch history information is not available.

We ran a microbenchmark to demystify the details of Intel's static branch prediction, which Intel does not disclose. Our microbenchmark consists of one test for every type of conditional branch (forward-taken, forward-not-taken, backward-taken, and backward-not-taken). For each type, we ran 1000 consecutive distinct branch instructions and counted the number of correctly predicted branches. Table~\ref{tab:branch} shows the result, averaged over 10 separate runs, on three CPU micro-architectures: Nehalem, Haswell, and Kaby-Lake.

\begin{table}[h!]
\caption{Static branch prediction rate for 1K branch instructions, each executed once}
\label{tab:branch}
\centering
\begin{tabular}{|c|c|c|c|}
\hline
branch type & Kaby-Lake & Haswell & Nehalem \\
\hline
backward taken & 76.3\% & 2.1\% & 99.9\% \\
backward not-taken & 96.1\% & 99.8\% & 0\% \\
forward taken &  0.6\% & 15.8\% & 0\% \\ 
forward not-taken & 95.3\% & 99.1\% & 99.3\% \\
\hline
\end{tabular}
\end{table}

The results suggest that while the old processor (Nehalem) used a direction-based branch prediction, the newer processors (Haswell and Kaby-Lake) predict most branches as not-taken: a prediction strategy which favors spatial locality. In particular, backward branches usually belong to loops, and therefore, are more frequently taken. But Haswell often statically predicts them as not-taken. This strategy can potentially hurt branch prediction, but has the advantage that with good code locality, the wrong prediction path (not-taken) is fall-through and incurs a smaller penalty. Overall, for Haswell, code layout does not influence the branch prediction rate but better code locality can result in lower penalties.

Interestingly, the prediction rate of backward branches is significantly improved in Kaby-Lake (from 2.1\% in Haswell to 76.3\%) at the cost of a slight degradation in the prediction rate of the other three branch types (total of 22.7\% reduction in prediction rate).

For direction-based branch prediction (as used in Nehalem), code layout can influence the branch prediction rate. For instance, frequently taken branches can enjoy a higher prediction rate if they face backward. Similarly, infrequently taken branches are better predicted if they face forward. On the other hand, as we argued above, static branch prediction is of lower priority than locality. Therefore, we must ensure that imposing such orderings between branch instructions and their targets does not limit locality opportunities.

PH imposes these orderings after BB chaining and before function reordering, by joining the basic block chains of every function in a specific direction. This can potentially result in joining code segments which are not heavily related. Furthermore, the larger code segments can limit the benefit in function reordering. Instead, here we demonstrate that Codestitcher can impose such orderings without unnecessarily joining basic blocks together. 

To this end, we introduce a new stage between basic block chaining and code collocation, called \emph{BB chain partial ordering}. The role of this stage is to identify an ordering among basic block chains which is optimal for branch prediction. The code collocation stages then explore locality opportunities within this partial ordering. 

BB chain partial ordering uses a framework similar to code collocation. First, it defines a directed weighted graph on the instruction sequences. The weight of each edge $(S,T)$ indicates the branch prediction benefit when $S$ precedes $T$ in the final layout. We explain the precise formulation of edge weights in the next section. Codestitcher iteratively chooses the heaviest weight in the graph and sets the ordering between its corresponding BB sequences. Then, it updates the orderings between other BB sequences if their ordering is implied by transitivity via the newly inserted order. 

\subsubsection{Setting the Edge Weights for Branch Prediction}
For each branch instruction $b$, we define $\textit{BD}(b)$, the branch divergence of $b$, as the difference between the frequency count of $b$'s branch edges. More formally, let $t_b$ and $f_b$ respectively be the more likely and the less likely target of this branch. Then
\begin{equation}
	\textit{BD}(b) = F(b,t_b) - F(b,f_b).
\end{equation}
This value indicates how far the branch targets are in terms of execution frequencies.
Now for two instruction sequences $S$ and $T$, we define the branch prediction profit of $S$ preceding $T$ as 
\begin{align*} 
	\textit{BPP}(S,T)  & = \sum_{\substack{\textit{b is a branch}\\ b \in S\\ f_b \in T}} \textit{BD}(b) 
 + \sum_{\substack{\textit{b is a branch}\\ b \in T \\ t_b \in S}} 
\textit{BD}(b) \\ & - \sum_{\substack{\textit{b is a branch}\\ b \in S\\ t_b \in T}} \textit{BD}(b) - \sum_{\substack{\textit{b is a branch}\\ b \in T\\ f_b \in S}} \textit{BD}(b).
\end{align*}

\section{Implementation}
\label{section:impl}

We implement Codestitcher using LLVM~\cite{LLVM:CGO04} version 3.9.  It can optimize x86 binaries (supporting other platforms would be straightforward) and supports multi-threaded code and multi-process applications.  Shared libraries that are profiled and compiled with Codestitcher are optimized as well.  Our implementation includes a compiler and a runtime profiler library. As a profile-guided optimization tool, Codestitcher has three stages: compiling the program to generate the profiler, running the profiler, and compiling the program to generate the optimized program.

\subsection{Instrumentation}

To generate the profiler, the user passes a flag to the compiler instructing it to instrument every shared library and executable in the program build tree by inserting calls to the profiler library. Codestitcher instruments the code at three places: function entries, basic block entries, and after every call site.  We implemented instrumentation as a link time optimization pass that transforms the x86 machine code generated by the compiler during code generation. The compiler assigns an 8-byte unique identifier to every basic block. The first two bytes identify the shared library or the executable, based on the hash value of its name. The next four bytes specify the function number, and the last two bytes specify the BB number.


\subsection{Profiler}
\label{sec:prof}
We implemented our profiler to perform edge-profiling.  Profiling basic block frequencies (node-profiling) is more efficient but requires static inter-procedural call graph analysis to estimate the edge frequencies from node frequencies. 
We opted to use edge-profiling as static inter-procedural call graph analysis is complicated by the presence of function pointers.

\begin{sloppypar}
The profiler performs edge-profiling via three main functions: \texttt{record\_function\_entry}, \texttt{record\_bb\_entry}, and \texttt{record\_callsite}. At every point, the profiler stores the previously executed basic block (\texttt{prev\_BB}). Upon executing \texttt{record\_function\_entry(BB)}, the profiler increments the call frequency between \texttt{prev\_BB} (caller) and \texttt{BB} (callee). The function \texttt{record\_bb\_entry(BB)} increments the jump frequency between \texttt{prev\_BB} and \texttt{BB}. 
The function \texttt{record\_callsite} updates \texttt{prev\_BB}, but does not update any frequency. To support multi-threaded programs, our profiler's run-time library uses thread-local storage to maintain the edge-frequencies on a per-thread basis.  Every thread emits its profiles into a separate file upon its termination. After the profiling run, profiles from different threads and processes are combined to form a single profile for the program.
\end{sloppypar}

\subsection{Layout Construction And Code Reordering}


The runtime profiles along with static CFG information are passed on to the layout construction module. The layout construction module also gets as input a list of distance levels. At each distance level $d$, the layout construction algorithm solves the $d$-close partial layout problem for every shared library and executable, and passes the generated partial layout on to the next distance level. 

At the end of this stage, Codestitcher will have generated a partial layout of all the hot code, consisting of instruction sequences which are not related enough to be joined together. Codestitcher generates a full layout by sorting these instruction sequences in decreasing order of their execution density. The execution density of an instruction sequence $S$ is defined as $\frac{\sum_{i \in S} f(i)} {size(S)}$, where $f(i)$ is the number of times the instruction $i$ has executed and $size(S)$ is the total size of $S$ in bytes. This allows hotter instruction sequences to appear closer to the beginning of the text section in the executable.

\section{Experimental Results}
\label{section:results}


In this section, we evaluate Codestitcher within two testing frameworks. First, we evaluate the instrumentation-based version of Codestitcher (as we explained above) on a machine with Haswell processors. Then we present an alternative implementation of Codestitcher, based on Linux perf tool, which does not require instrumentation. We compare the perf-based Codestitcher with LLVM's default profile-guided optimization framework (PGO), and on a machine with Kaby-Lake processors. Then, we discuss the overheads of both Codestitcher versions and compare them against LLVM's PGO.

\subsection{Experimental Setup}
For the instrumentation-based Codestitcher, we use a test machine which runs Ubuntu 16.04 and is powered by two dual core i5-4278U Haswell processors running at 2.60~GHz.
Using the {\tt getconf} command, we determined that the processor has a 32~KB instruction cache and a 256~KB second level unified cache private to each core. Both caches are 8-way set associative. The last level cache is 3~MB, 12-way set associative, and is shared by all cores.  We compile all our benchmark programs with Clang~3.9 
with second optimization level (O2) and link time optimizations enabled with the \texttt{-flto} flag.

\subsection{Benchmarks}
\label{sec:benchmarks}

We test a set of widely-used programs that have large code sizes: a database server (MySQL cluster 7.4.12), a browser (Firefox 50.0), a compiler (Clang 3.9), a complete Apache web server (httpd-2.4 along with PHP-7.0), and a python interpreters (Python 2.7.15). In this section, we first discuss the code size and cache characteristics of these programs and then describe the testing setup for each.

Our test suite includes a total of 7 tests over these 5 applications, as shown in Table~\ref{tab:char}. The programs contain between 15~MB (MySQL) and 81~MB (Firefox) in the \texttt{text} section of their executables, except for Python whose text size is 2~MB (all code sizes are rounded up to 1~MB).
Two programs, Clang and Apache, occupy multiple rows in the table, because they are tested on two configurations. Other programs Firefox and Python have multiple binaries.  The table shows the largest one.

\begin{table}[h!]
\caption{Code size and performance characteristics (MPKI means misses per thousand instructions)}
\label{tab:char}
\bgroup
\def\arraystretch{1.1}
\setlength\tabcolsep{4pt} 
\begin{tabular}{|c|c|c|c|c|}
\hline
\shortstack{applications \&\\tests} &\shortstack{\vspace{1mm}\\I-Cache\\MPKI} & \shortstack{I-TLB\\MPKI} & \shortstack{binaries \& their\\.text size(MB)}  \\
\hline
MySQL &  62.44 & 9.35  & mysqld(15)    \\ \hline
\begin{tabular}{l|l}\multirow{2}{*}{Clang}&-j4 \\ &-j1\\ \end{tabular} &  \begin{tabular}{c} \multirow{2}{*}{} 14.40\\ 8.14\end{tabular} & \begin{tabular}{c} \multirow{2}{*}{} 2.23\\1.01 \end{tabular} & 
clang(50), gold(4) \\ \hline
Firefox & 9.16 & 1.54 & libxul.so(81)  \\ \hline
\begin{tabular}{c|c}\multirow{2}{*}{\rotatebox[origin=c]{90}{Apache}}&w/ opcache \\ &w/o opcache\\ \end{tabular}
&   \begin{tabular}{c} \multirow{2}{*}{} 4.13\\7.63 \end{tabular} 
& \begin{tabular}{c} \multirow{2}{*}{} 0.33\\0.96 \end{tabular} 
& \begin{tabular}{c} \multirow{3}{*}{} libphp7.so(14) \\ 
									  httpd(1) \\ 
                                     opcache.so(1) 
                                     
         \end{tabular} \\ \hline

Python & 3.40 & 0.19 & python(2) \\ \hline

\end{tabular}
\setlength\tabcolsep{6pt} 
\egroup
\end{table}

We use measure the I-Cache and I-TLB misses, because their frequent misses lead to frequent stalls at the CPU front-end and limit the ability of out-of-order execution.~\footnote{We use the ICACHE\_64B:IFTAG\_MISS and ITLB\_MISSES:MISS\_CAUSES\_A\_WALK event hardware performance counter events to measure I-Cache and I-TLB misses.} The applications in Table~\ref{tab:char} are ordered by their I-Cache MPKI (misses per thousand instructions), from the highest, 62 MPKI for MySQL, to the lowest, 3.4 MPKI for Python.  The cache performance does not always correlate with code size directly.  Firefox, the largest program on our benchmark, trails Clang-j4, and MySQL in I-Cache miss rates.      

The instruction cache performance correlates well with that of the instruction TLB, shown by the I-TLB MPKI column.  The TLB performance is identically ordered among the applications from 9 MPKI for MySQL to 0.2 MPKI for Python. 

The cache and TLB performance depend on the size and usage of the executed code, which change in different applications and in different runs of the same application.  For Clang, the miss rates almost double when moving from the \texttt{-j1} test to the \texttt{-j4} test.  Another important factor is the degree of multi-threading. MySQL and two Apache tests are multi-threaded, and more threads results in higher pressure on the I-Cache and I-TLB.\@

Although all applications suffer from poor instruction performance, there is a large variation among them.  The highest and lowest MPKIs differ by a factor of 18 times for the instruction cache and 49 times for TLB.\@ 


In a recent study~\cite{Ottoni:CGO17}, Ottoni and Maher measured the instruction cache performance of 4 server applications used internally at Facebook, including the execution engine that runs the Facebook web site~\cite{Ottoni:CGO17}.  These commercial applications are not available for our study.  When we downloaded the open-source version of one of their applications, HHVM, we found that the program text after compilation is 24~MB, much lower than 133~MB, as reported by them. The baseline compilation of HHVM with Clang resulted in a broken binary. Therefore, we were unable to test this program.

Compared to our test suite, these commercial applications have larger code sizes, between 70~MB and 199~MB (compared to 2~MB and 81~MB among our applications).  They show similar ranges of I-Cache MPKIs, from 5.3 to 67 (compared to 3.4 to 62), and I-TLB MPKIs, from 0.4 to 3.1 (compared to 0.19 to 9.4), as the programs in our test suite do.  Our MySQL test is an outlier with its 9.4 I-TLB MPKI, 3 times the highest number in the Facebook study. 

Next we describe the testing setup for each program.

\paragraph{Python} 
For testing the Python interpreter, we use the \emph{apps} benchmark group from \emph{Google's Unladen Swallow Benchmark Suite}~\cite{UnladenSwallow} as input scripts.
The \emph{apps} benchmark group consists of 6 ``high level'' applicative benchmarks: \emph{2to3}, \emph{chameleon\_v2}, \emph{html5lib}, \emph{rietveld}, \emph{spambayes}, and \emph{tornado\_http}. The Unladen swallow benchmark provides a test harness which reports the average running time along with the standard deviation. It also provides three options for the runtime duration: \emph{fast}, \emph{regular}, and \emph{rigorous}. We use \emph{fast} for profiling and \emph{regular} for testing. We report the improvement over total wall-clock runtime of the 6 inputs.
 
\paragraph{Firefox}
For Firefox, we use the \emph{tp5o} benchmark from Talos~\cite{Talos}, a performance testing framework used at Mozilla. The \emph{tp5o} benchmark tests the time it takes for Firefox to load 49 common web pages stored on the local file system. For each web page, Talos measures its load for 25 times, discards the first 5, and takes the median of the remaining 20. It then reports a single score as the geometric mean of these numbers over all web pages. 



\paragraph{MySQL}
For testing the MySQL server, we use the non-transactional read-only test (\emph{oltp\_simple}) from the Sysbench benchmark suite. We run both the Sysbench client and the MySQL server on the same machine. For both profiling and testing, we use 4 client threads to read from a database pre-populated by 8 tables each with 100,000 records. Sysbench reports the total throughput (requests per second) along with the average latency and tail latency (95 percentile) of requests. We report the mean improvement in average latency over 10 runs.


\paragraph{Apache}
Our setup uses the Apache web server along with the PHP7 interpreter module, the PHP bytecode cache module (OPCache), and the MYSQL database server; we therefore have Codestitcher optimize all four executables. We set up an isolated network between a single client and a single server using a 1~Gbps Ethernet cable. Our test data is WP-Test~\cite{WPTest}, an exhaustive set of WordPress~\cite{Wordpress} test data. We test the server with OPCache both being enabled and disabled. For testing, we run Apachebench on the client side to make a total of 10,000 requests to the server over 50 concurrent connections.
Apachebench reports the throughput along with service time quantiles. We report the 90th percentile of service times.

\paragraph{Clang}
Our Clang experiment optimizes both the Clang compiler and the Gold linker. We test Clang by compiling the applications in LLVM's MultiSource test suite, which consists of 25 applications including sqlite3. For each compilation, the test suite reports separate user times of compilation and linking. We measure the total user time.  We perform two separate experiments: in one experiment, we run \texttt{make -j 4} to build all applications using 4 parallel jobs. In the other, we use a single compilation job (\texttt{make -j 1}). We repeat each experiment 10 times, aggregate the user times of compilation and linking for each application, and report the total user time of compilation over all 25 applications.

\subsection{Comparison}
\label{sec:compare}

\begin{figure*}[tb]
\centering
\includegraphics[width=0.99\linewidth]{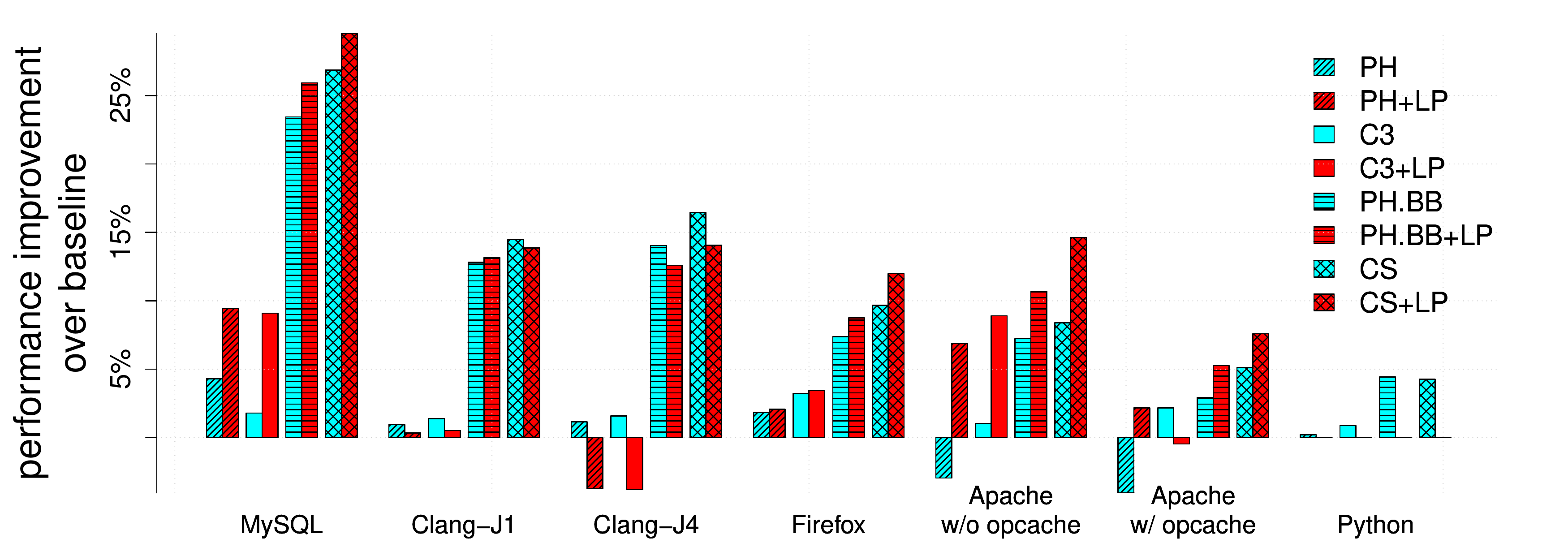}
\caption{Performance improvement over baseline}
\label{figure:results}
\end{figure*}




We compare the improvements from Codestitcher (CS) against three code layout techniques.

\begin{itemize}
\item {\textbf{PH.BB}}: Pettis-Hansen's full basic block reordering technique, consisting of function splitting, basic block chaining, and function reordering, implemented by us.
\item {\textbf{PH}}: Pettis-Hansen's function reordering adopted from HHVM~\cite{HHVM} (see the last paragraph of this section).\@
\item {\textbf{C3}}: Function placement based on chain clustering as proposed by Ottoni and Maher~\cite{Ottoni:CGO17}, adopted from HHVM.\@
\end{itemize}

We also evaluate each of the four techniques (PH, C3, PH.BB, and CS) when using large pages for code (resulting in the new techniques PH+LP, C3+LP, PH.BB+LP, and CS+LP). Our implementation of large pages uses anonymous large pages for the hot code, similar to that of Ottoni and Maher~\cite{Ottoni:CGO17}.



C3 and PH have both been implemented as part of HHVM, an open-source PHP virtual machine, developed at Facebook. 
The two function reordering heuristics rely on the Linux {\bf perf} utility to collect profile data. The {\bf perf} tool interrupts the processes at the specified frequency and records the stack trace into a file.  After the profiling run, hfsort reads this profile, builds the function call-graph, and computes the function layouts for each heuristic.

\subsection{Profiling}
For function reordering methods (C3 and PH), we use the \emph{instructions} hardware event to sample stack traces at the sampling rate of 6250 samples per second. For each program, we profile over 5 iterations of our test run. 
Additionally, for each application, we decompose its profile into profiles for each of its constituent binaries (those listed in Table~\ref{tab:char}). This allows us to separately optimize each executable and shared library in the program.


For Codestitcher and PH.BB, we use the profiles generated by our own profiler described in Section~\ref{sec:prof}. Since our profiler collects the full trace, we profile over shorter runs. 
For Clang, we compile the multiSource benchmark only once. For Firefox, we load every other web page from the list of 49 web pages in \emph{t5po}. For Python, we profile over the ``fast'' run of the inputs scripts. For Apache web server, we use a lighter load of 100 requests over 4 concurrent connections. We perform separate profiling runs for each OPCache configurations (enabled and disabled) and combine the two profiles. The MySQL tests are controlled by the execution time. Therefore, we use the same configuration as the test run.

\subsection{Results}
\label{sec:results}

\begin{figure*}[tb]
\includegraphics[width=0.99\linewidth]{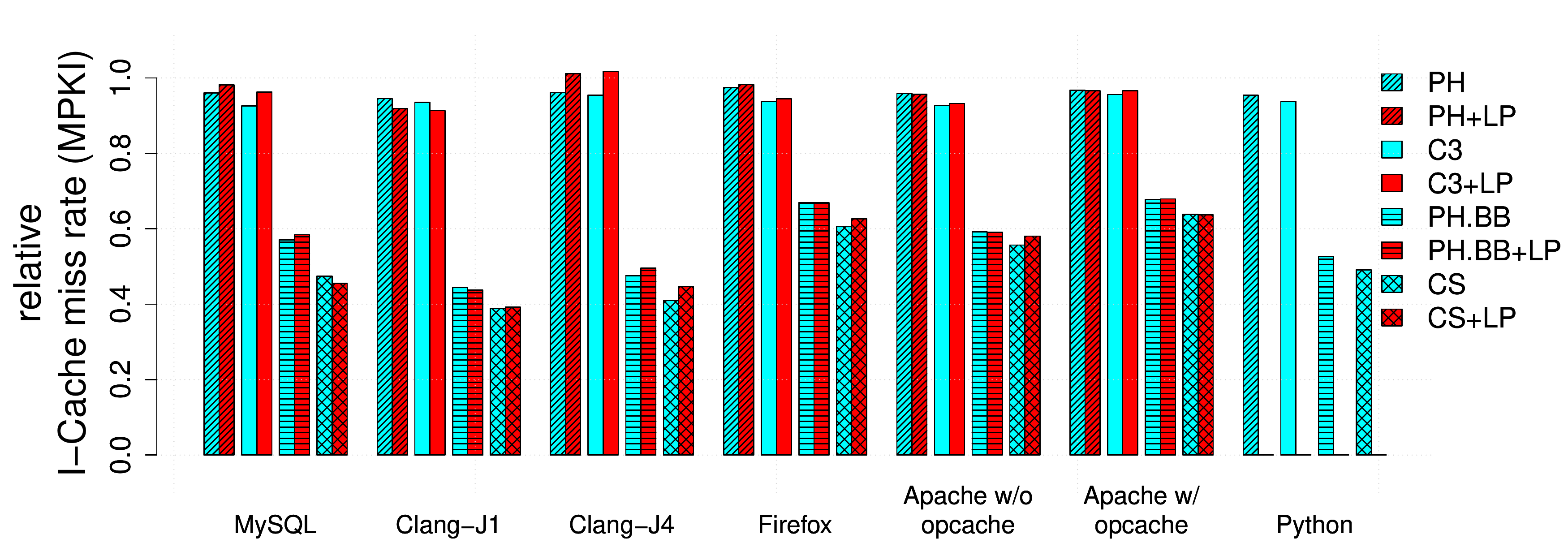}
\caption{L1 instruction cache miss per 1K instructions, relative to the baseline.}
\label{figure:results-CMPKI}
\includegraphics[width=0.99\linewidth]{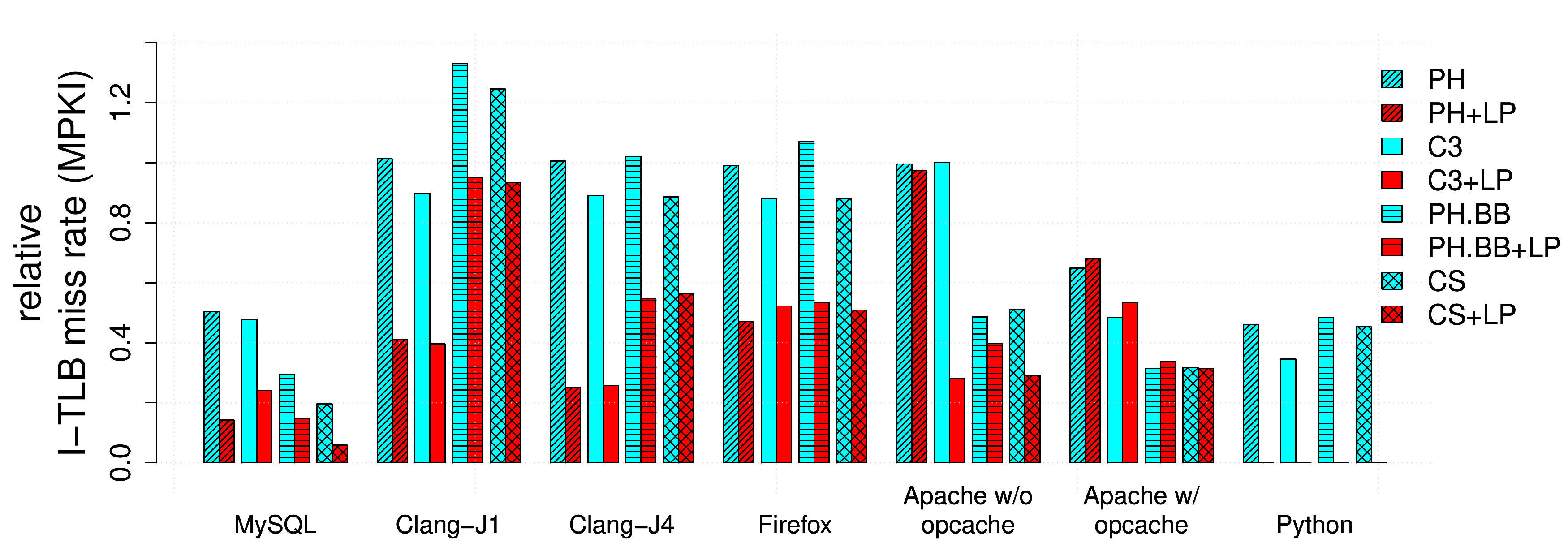}
\caption{L1 instruction TLB miss per 1K instructions, relative to the baseline.}
\label{figure:results-TMPKI}
\end{figure*}

Figure~\ref{figure:results} shows the performance improvement by 8 optimization techniques on all 7 tests. The programs are shown from left to right in decreasing order of I-Cache and I-TLB MPKI (Table~\ref{tab:char}). Python is only tested with regular pages because its hot code does not exceed a single MB.\@ Therefore, no large pages will be used and the large page (LP) results will be the same as using regular pages.

It is evident that the higher I-Cache and I-TLB pressure (from left to right) a program exhibits, the more effective code layout optimization is, and the greater improvements we see in performance. 

The speedups from CS range between 5.1\% (for Apache with opcache) and 26.9\% (for MySQL), with about half of the tests (3 tests) gaining more than 10\% speedup. CS+LP  delivers slightly higher improvements: from 5.0\% for Python to 29.5\% for MySQL. It improves 5 tests by more than 10\% (all tests except Apache with opcache and Python).

\begin{table}[t]
\caption{Geometric mean improvement across all tests (replacing LP results for Python with the regular page results)}
\label{tab:geomean}
\centering
\begin{tabular}{c|c|c|c|c}
& PH & C3 & PH.BB &  CS \\ \hline
regular pages & 0.2\% & 1.7\% & 10.1\% & \bf{11.9\%} \\ \hline
large pages & 2.4\% & 2.5\% & 11.4\% & \bf{13.5\%}  \\ \hline
\end{tabular}
\end{table}

PH.BB is the best previous solution among these tests. CS outperforms PH.BB in all tests except when using regular pages for Python. Table~\ref{tab:geomean} shows the geometric mean improvements across all tests, for each of the 8 optimization techniques (We replace the LP results for Python with the regular page results). We observe that CS dominates other techniques both when using regular pages and large pages. On the other hand, if we take PH.BB as the baseline technique, CS and CS+LP improve over PH.BB and PH.BB+LP, respectively by 1.6\% and 1.9\%.

\begin{table}[b]
\caption{Geometric mean of relative L1 instruction cache, instruction TLB, L2, and LL cache MPKI, and branch misprediction rate, across all tests (replacing the large page results for Python with regular page results)}
\label{tab:geomean-perf}
\begin{tabular}{c|c|c|c|c}
& PH.BB & PH.BB+LP & CS & CS+LP \\ \hline
\vspace{1mm}
relative L1-cache MPKI & 55.7\% & 55.9\% & \bf{50.6\%} & 52.1\% \\ \hline
relative I-TLB MPKI & 69.1\% & 51.3\% & 64.5\% & \bf{47.2\%} \\ \hline
\shortstack{relative branch misprediction rate} & 95.4\% & \bf{94.9\%} & 97.6\% & 97.5\% \\ \hline
relative L2-cache MPKI & 77.0\% & 77.6\% & \bf{74.2\%} & 75.0\%  \\ \hline
relative last level cache MPKI & 87.3\% & 90.8\% & \bf{86.4\%} & 88.0\%  
\end{tabular}
\end{table}

The relative I-Cache miss rates are shown in Figure~\ref{figure:results-CMPKI}.

The improvement in I-Cache miss rate is mostly governed by the code layout granularity (basic block ordering vs. function ordering). However, CS still dominates PH.BB, in each of the 7 tests.

We also observe that regardless of the code layout technique being used, using large pages rarely leads to additional improvement of the I-Cache miss rate. Especially for Clang-J4, using large pages is detrimental for all techniques. Between PH.BB and CS, the highest win from large pages is observed when applying CS on MySQL (2.1\%). For PH.BB, the maximum improvement from large pages is 0.7\% and happens for Clang-J1.

Figure~\ref{figure:results-TMPKI} shows the I-TLB miss rates relative to the baseline.

When using regular pages, ordering at basic block granularity usually reduces the I-TLB miss rates compared to function granularity. Significant differences can be observed for four tests (MySQL, Clang-J1, and Apache, with and without opcache) among which basic block ordering wins in three. 

With large pages, the effect is mixed: basic block ordering results in significant degradation for two tests (Clang-J1, and Clang-J4) while it improves Apache with opcache, and delivers comparable results in the other three. We believe that mispredicted branches are to be blamed for this phenomenon. Specifically, in a basic block layout, where function fragments are stored far from each other, speculative long jumps across distant function fragments may result in inferior TLB-performance. Notably, this does not happen in a function layout where every function is stored contiguously in memory.

On the other hand, we observe that unlike I-Cache miss rate, for every individual code layout technique, using large pages has a consistent positive effect on the I-TLB miss rate.

For an overall comparison, we report in Table~\ref{tab:geomean-perf}, the geometric mean of I-Cache and I-TLB miss rates across all 8 tests, along with those of L2 and LL cache, and branch prediction. On average, CS gives the lowest miss rates on all three cache levels. CS+LP is the winner in I-TLB performance, beating CS by an absolute difference of 17.3\%. Considering that CS+LP gives 1.6\% higher overall improvement than CS (Table~\ref{tab:geomean}), we infer that its superior I-TLB performance dominates the effect of its slightly higher cache miss rates. Branch prediction is the single weakness of CS, where it trails PH.BB by at most 2.6\%.


Finally, for function reordering, C3 outperforms PH in all tests except one. The largest differences can be observed in Firefox (1.4\%) and Apache (up to 6\%). For Apache, PH with regular pages always degrades the performance, while C3+LP results in 9\% improvement for Apache without opcache. On average, C3 improves over PH by a difference of 1.9\%. Using large pages reduces this difference to 1\%.



\subsection{Codestitcher Without Instrumentation and Comparison with PGO}
\label{sec:cs-perf}
In this section we present and evaluate our alternative implementation of Codestitcher which does not require instrumentation. During compilation, we emit unique symbols at the beginning of every basic block. Then we leverage the Linux \textbf{perf} utility to gather profiles by sampling the \emph{last branch record} (LBR) stack. As the name suggests, the LBR stack records the most recent instructions which have caused a control transfer (conditional branches, jumps, calls, and returns). Thus it includes all control transfers between basic blocks except fall-throughs. 

Fall-throughs can be inferred from the LBR stack as follows. For every two consecutive branches in the LBR stack, all instructions between the target address of the first branch and the source address of the second branch are executed contiguously, as no other branches can execute in between. These contiguous instruction sequences include all fall-throughs between basic blocks.

We use the generated profiles to compute the sampled weighted inter-procedural control flow graph. Then we apply the hierarchical code merging algorithm as we explained in Section~\ref{sec:hier}, obtain the optimal code layout and recompile the program according to the new layout.

\subsubsection{Comparison against PGO}
Now we evaluate our perf-based Codestitcher on a more powerful hardware platform and compare it against LLVM's default profile-guided optimization (PGO) framework.

The new hardware platform runs Ubuntu 16.04 and is powered by two quad-core i7-7700 Kaby-Lake processors running at 3.60GHz. Similar to the Haswell micro-architecture, each core has a 32~KB L1 instruction cache and a 256~KB L2 unified cache, each with associativity of 8. The shared last level cache is 8~MB and 16-way set associative. The instruction TLB includes 64 entries for regular pages, with associativity of 8, and 8 fully-associative entries for large pages. The shared L2 TLB contains 1536 with associativity of 6.

LLVM's PGO works in multiple stages. First, the compiler instruments the program at IR level for edge-profiling. When the instrumented program runs, it generates basic block execution count profiles. In the third stage, these profiles are fed into another compilation phase to guide intra-function basic block placement and inlining heuristics.

First, we use the PGO-instrumented program to perform a single run of each profiling input and then use the emitted profiles to build the PGO-optimized programs.

Since LLVM's PGO optimization does not perform function reordering, we complement its affect by subsequently applying each of the two function reordering techniques in Section~\ref{sec:compare} (C3 and PH) on the PGO-optimized programs. To do so, we perform a second stage of profiling on PGO-optimized programs using HHVM's perf-based call-graph profiling infrastructure (as we described in Section~\ref{sec:compare}). The obtained profiles are then used to compute the new PH and C3 function orderings and rebuild the programs with such orderings. Effectively, these optimized programs achieve intra-function basic-block reordering via PGO and function reordering via PH and C3.

In a third stage, separate from the prior stages, we profile the baseline program using our own perf-based profiling framework (as we described in Section~\ref{sec:cs-perf}) to generate inter-procedural basic block level profiles. Codestitcher then uses these profiles within its hierarchical framework (as we described in Section~\ref{sec:hier}) to construct its optimal inter-procedural basic block layout. Furthermore, we apply our reimplementation of Pettis-Hansen's full basic block reordering (PH.BB) as we described in Section~\ref{sec:compare}, on the same profiles. 

In summary, we evaluate the following code layout techniques.

\begin{itemize}
\item PGO: PGO-optimized program
\item PGO.PH: Pettis-Hansen function reordering applied on top of PGO
\item PGO.C3: Call-Chain-Clustering function reordering applied on top of PGO
\item PH.BB: Global Pettis-Hansen basic block reordering using our perf-based profiling framework, applied on top of baseline
\item CS: Codestitcher using our perf-based profiling framework, applied on top of baseline
\end{itemize}

Additionally, we evaluate each of the code layout techniques (except PGO) when large pages are used for hot code.

Our experiment features five tests: MySQL, Clang-J1, Firefox, Apache without opcache, and Python. Our experimental setup is slightly different from Section~\ref{sec:benchmarks}, as we discuss next.

Most importantly, we test every application on a completely disjoint set of profiling and testing inputs: for MySQL we use the combination of \emph{select}, \emph{select\_random\_points}, \emph{update\_index}, and \emph{insert} from sysbench for profiling and \emph{oltp\_simple} for testing. For Clang, we use \emph{multiSource applications} for profiling and \emph{multiSource benchmarks} for testing, and only test with one compilation job (-j1). For Firefox, we use a combination of tests from Talos (a11yr, tsvgx, tscrollx, tart, cart, kraken, ts\_paint, tpaint, tresize, and sessionrestore) for profiling and tp5o for testing. For Apache, we use a clean installation of Drupal~\cite{Drupal} for profiling and WP-Test~\cite{WPTest} for testing, and only test with OPCache disabled. In addition, and in contrast to the Apache tests in Section~\ref{sec:benchmarks}, we exclude the optimization of MySQL from the Apache test. Finally, for Python, we profile using the combination of \emph{etree} and \emph{template} scripts from the unladen-swallow benchmark suite. For testing, we use the \emph{apps} scripts, as before.

For each application, we use the same profiling inputs across all three profiling stages. We generate profiles over a single run for the full-trace PGO profiling, and over five identical runs for the other two perf-sampling-based profiling stages. For testing each technique, we report the average improvement over 10 identical runs of the test input, relative to the baseline program. Here, we also report the standard deviation of the improvement. 

We apply Codestitcher on two layout distance parameter: 4~KB and 1~MB. The 4~KB parameter is the most important and will include most of the control transfers in the optimized code. The 1~MB distance parameter intends to reduce the interference at L1 and shared TLBs besides reducing internal fragmentation for large pages.

Finally, all programs are compiled using the ``-flto -O3'' flags.

\begin{figure*}[tb]
\centering
\includegraphics[width=0.99\linewidth]{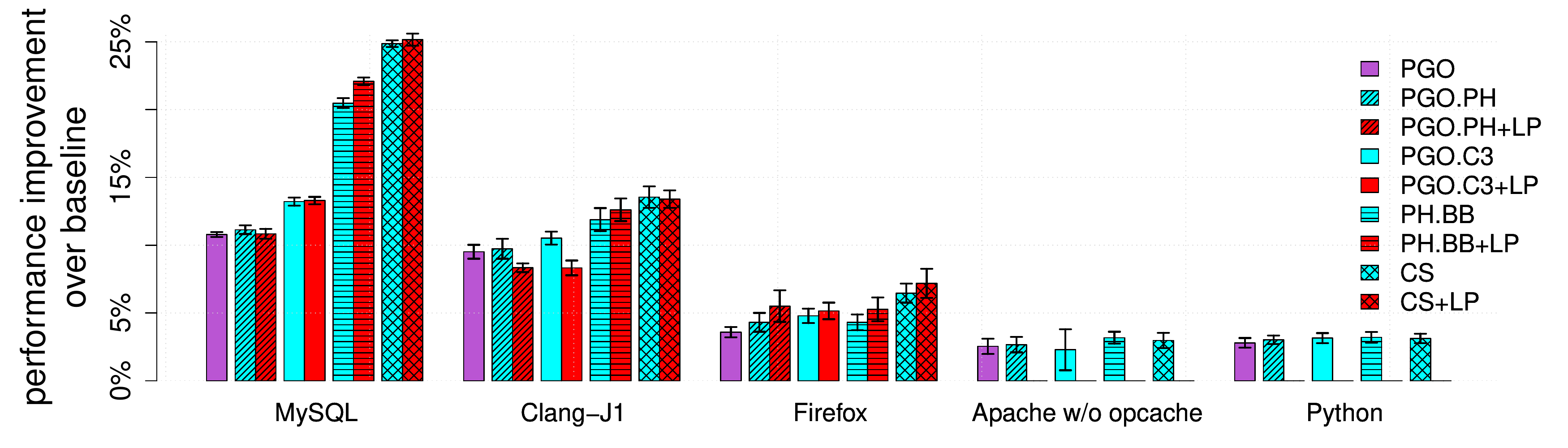}
\caption{Comparison of PGO, PGO.PH, and PGO.C3 against our perf-based techniques (BB.PH and CS) in terms of performance improvement over the baseline}
\label{fig:results-pgo}
\end{figure*}

Figure~\ref{fig:results-pgo} shows the results. For Python and Apache, the hot code in each of their executing binaries will not exceed a single MB.\@ Therfore, we do not evaluate the techniques with large pages on these two programs. The other three applications are optimized using each of the 9 optimization techniques as we described above.

We observe that CS significantly outperforms the other techniques in the largest three applications (MySQL, Clang, and Firefox). All techniques perform similarly well on Apache and Python. 

We observe that applying function reordering on top of PGO results in modest improvements. The highest improvements over PGO by function reordering techniques are 2.6\% on MySQL (by PGO.C3+LP), 1\% on Clang (by PGO.C3), and  2\% on Firefox (by PGO.PH).

PH.BB is the second best technique, after CS.\@ The larges difference between CS and PH.BB is seen in MySQL (almost 5\%). For Firefox, PH performs similiarly well as the function reordering techniques. For smaller programs (Apache and Python) PH performs marginally better than the other techniques (outperforms CS by 0.2\% and 0.8\%, respectively). 

We also observe that overall, the use of large pages does not lead to a significant improvement. In particular for Clang, using large pages has a negative impact on both function reordering techniques, resulting, respectively, in 1.5\% and 2\% reduction in the improvements from PGO.PH and PGO.C3.

For an overall comparison, Table~\ref{tab:geomean-pgo} reports the geometric mean improvements of all 9 techniques on the 5 tests.

\begin{table}[t]
\caption{Geometric mean improvement across all 5 tests in Figure~\ref{fig:results-pgo} (Large page results for Python and Apache are replaced with the regular page results)}
\label{tab:geomean-pgo}
\centering
\begin{tabular}{c|c|c|c|c|c}
& PGO & PGO.PH & PGO.C3 & PH.BB &  CS \\ \hline
regular pages & 5.8 & 6.1\% & 6.7\% & 8.4\% & \bf{9.9\%} \\ \hline
large pages & NA & 6.0\% & 6.4 \% & 9.0\% & \bf{10.1\%}  \\ \hline
\end{tabular}
\end{table}

\subsection{Overhead}
Each of the code layout optimization techniques have their own costs and overheads. These overheads are categorized into three types, each corresponding to one stage of the profile-guided optimization framework: 
\begin{itemize}
\item Profiling overhead: slowdown of the program due to profiling
\item Trace processing overhead: additional processing time to compute the total node/edge counts from emitted profiles
\item Layout construction and code reordering overhead: additional processing and build time to compute the optimal layout and reorder the program's binary according to that layout
\end{itemize}

We report the profiling overhead for both the instrumentation-based the perf-based Codestitcher frameworks, along with LLVM's PGO.\@ In addition, we report the trace processing overhead and the build overhead for the perf-based technique. 

The overhead results are shown in Table~\ref{tab:overhead}.

The profiling overheads are measured differently, and in accordance with our experimental setup: 90\% tail latency increase in Apache, average latency in MySQL, increase of total compilation and link time for Clang, increase of the score reported by the \emph{talos} benchmark suite for Firefox, and increase of total wall-clock time for Python.

The trace processing overheads for (perf-based) Codestitcher are reported as the processing time relative to the length of the profiling run. This processing reads all the sampled LBR stack traces and builds the weighted inter-procedural CFG.\@ We have implemented this inside the Linux perf utility.

The layout construction and link overheads are reported as the excess build time relative to the link time of the programs (given all the LLVM bitcode files). For the perf-based Codestitcher, our layout construction library is prototypical and is implemented outside the compiler and in Ruby.

\def\arraystretch{1.5}%
\begin{table}[h!]
\caption{Overheads relative to baseline program, for three optimization frameworks: Codestitcher with instrumentation, perf-based Codestitcher, and LLVM's PGO.\@}
\label{tab:overhead}
\centering
\begin{tabular}{|c|c|c|c|c|c|c|} 
\cline{3-7}
\multicolumn{2}{c|}{} & MySQL & Clang & Firefox & \shortstack{\\Apache w/o\\opcache}  &  Python\\
\specialrule{.15em}{0em}{0em} 
\hline
\shortstack{\\Codestitcher\\(with instrumentation)} & profiling & $14\times$ & $53\times$  & $64\times$ & $3\times$  & $62\times$ \\ 
\specialrule{.15em}{0em}{0em} 
\hline
\multirow{3}{*}{\shortstack{Codestitcher\\(perf-based)}} & profile sampling  & 6.7\% & 5.3\% &  3.7\%  & 7.2\% & 5.5\%  \\ \cline{2-7}
& \shortstack{trace processing} & 21\% & 10\%   & 24\% & 23\% & 6\% \\ \cline{2-7}
& \shortstack{\\layout construction\\and code reordering} & 329\% & 54\%   & 161\% &  91\%  & 456\% \\
\specialrule{.15em}{0em}{0em}
\hline
\shortstack{LLVM PGO} & profiling & 45\% & 149\% & 113\%  &  142\% &   4\% \\
\specialrule{.15em}{0em}{0em} 
\hline
\end{tabular} 
\end{table}
\def\arraystretch{1}%

The overhead results indicate that our perf-based Codestitcher incurs the least slowdown among the three techniques. Furthermore, it requires no instrumentation for profiling (except for emitting symbols in the program), and can be applied iteratively, while inflicting the least interference with the program execution. Two major drawbacks of the perf-based technique are the higher overhead of layout construction and its storage overhead for traces. Our prototypical layout construction library can be significantly be optimized by reimplementation in the compiler or in the perf tool. The storage overhead can be reduced by performing trace processing in parallel with trace collection.


\subsection{Evaluation on Nehalem}
\label{sec:nehalem}

\begin{figure}[b]
\includegraphics[width=0.6\linewidth]{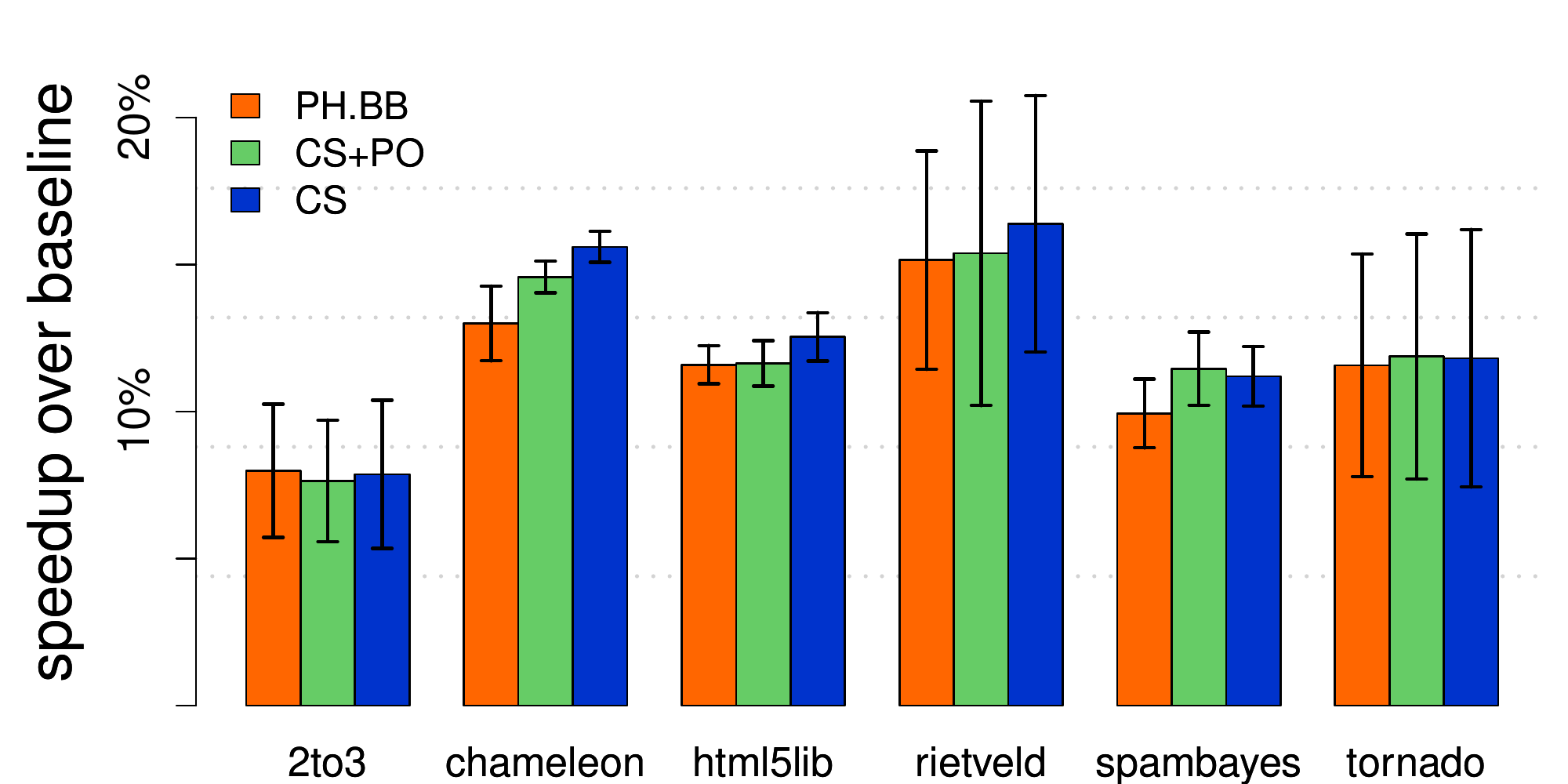}
\caption{Comparison between CS, CS with partial ordering (CS+PO) and PH.BB.\@ Evaluated on Nehalem processor}
\label{fig:nehalem}
\end{figure}

As we discussed in Section~\ref{sec:br}, Nehalem uses a direction-based branch prediction method, which is different from  Haswell's ``always-not-taken" method.  In this section, we evaluate the interaction of code layout optimizations with branch prediction.
As we explained in Section~\ref{sec:br}, Codestitcher can add a middle step between BB Chaining and code collocation, to jointly optimize branch prediction and locality, with minimal conflict on each other. The resulting new technique is called CS+PO.\@ Here, we run an experiment on Python, using the same test setup as we described in Section~\ref{sec:benchmarks}. We compare three techniques: PH.BB, CS, and CS+PO.\@
We measure the speedup, I-Cache and I-TLB miss rates, relative to the unoptimized Python. Table~\ref{tbl:misp} shows the geometric mean of these numbers across all six Python inputs. We also report the individual speedup for each input, along with the standard deviation in Figure~\ref{fig:nehalem}. The standard deviation results are shown here in order to distinguish between significant and non-significant improvements.



The comparison between PH.BB and CS+PO shows the effect of limited locality improvement in PH.BB, caused by its BB chaining strategy. CS shows the effect of unrestricted BB collocation.

It is clear that even on Nehalem, unlimited BB collocation is the best design.  It has a modest performance gain, but a clear improvement in cache and TLB performances. For the instruction cache, the relative MPKI for CS is 8\% lower than PH.BB and 4.4\% lower than CS+PO.\@

A greater effect is shown by the ITLB performance.  CS+PO causes 10 times as many ITLB misses as PH.BB, as a result of limited collocation.  Partial ordering apparently increases the page working set significantly.  With CS, however, not only is the ITLB problem of inter-procedural BB ordering  removed, CS actually improves.  By splitting a function into more pieces, CS reduces the ITLB MPKI by 16\% compared to the mere function splitting by PH.BB.\@

Finally, all techniques similarly reduce the branch misprediction rate: PH.BB by 11\% and CS and CS+PO both by 10\%.

The individual results (Figure~\ref{fig:nehalem}) are consistent with the average improvement results. CS gives highest improvement in all but two input scripts: \emph{2to3} (most improved by PH.BB) and \emph{spambayes} (most improved by CS+PO).
Especially, CS is the best technique on \emph{chameleon} and \emph{html5lib}, the longest running input scripts among the six. Furthermore, the error bar for CS has no overlap with others on \emph{chameleon} and has small overlap on \emph{html5lib}, which is an indication of its definitive improvement.

\begin{table}[t]
\caption{Comparison between PH.BB, Codestitcher with partial ordering, and just Codestitcher}
\label{tbl:misp}
\begin{tabular}{c|c|c|c}
& PH.BB & CS+PO & CS \\
\hline
speedup & 11.5\% & 12.1\% & \bf{12.6\%} \\
\hline
relative ICACHE MPKI & 39.4\% & 34.8\% & \bf{31.4\%} \\
\hline
relative ITLB MPKI & 2.6\% & 26.1\% & \bf{2.2\%} \\
\hline
relative branch misprediction rate & \bf{88.9\%} & 90.0\% & 90.1\% \\
\end{tabular}
\end{table}

\section{Related Work}
\label{sec:related}
Code layout is a form of cache-conscious data placement, which has been shown to be NP-hard, even for an approximate solution~\cite{PetrankR:POPL02,Lavaee:POPL16,AttiyaY:ICPDS17}. However, code layout is more tractable for two main reasons. First code access patterns can be precisely captured from program execution. Second, the less structured format of code allows code reordering to be done with higher flexibility.

Ramirez et al.~\cite{Ramirez:ISCA01} specifically studied code layout optimizations for transaction processing workloads. 
They implemented PH in the Spike binary optimizer~\cite{Cohn:DTJ98}, and studied the impact of different stages of PH.\@
Their implementation mostly follows PH, but uses a fine grain function splitting, analogous to us. Their work does not give insights into the implications of code layout for branch prediction. 

More recently, Ottoni and Maher~\cite{Ottoni:CGO17} introduced call-chain clustering, a new heuristic for function reordering.
Call-chain clustering follows a similiar bottom-up approach as PH,  but aims at reducing the expected call distance, by merging function layouts in the direct of the call. They evaluate their technique on four data-center applications including HHVM.\@ Our work gives an independent evaluation of their heuristic on programs which are more widespread, but have smaller code sizes.

Focusing on conflict misses, Gloy and Smith~\cite{GloyS:TOPLAS99} developed a solution based on a \emph{Temporal Relation Graph} (TRG). They defined the temporal relation between two code blocks as the number of times two successive occurrences of one block are interleaved by at least one occurrence of the other. The solution works best on direct-mapped caches. It looks at TRG edges running between code blocks mapped to the same cache lines. Considering these edges in sorted order, it tries to remove conflict misses by remapping procedures to other cache lines. Once all mappings are obtained, it orders the procedures to realize those mappings while minimizing the total gap between procedures.

The direct extension of TRG for $k$-way set-associative caches requires storing temporal relations between every code block and every $k$-sized set of code blocks. It is evident that as programs become larger and cache associativity goes higher, storing this information becomes more expensive. Liang and Mitra~\cite{LiangM:CASES10} introduced the \emph{intermediate blocks profile}, a compact model for storing this information, which enables them to evaluate the approach on up to 8KB associative caches, and on programs with up to about 500 procedures. Besides this, their optimizations are not portable across different cache configurations.

Finally, while in this paper we used edge-profiling to optimize code layout, path profiling~\cite{BL:MICRO96} gives more precise information about the control flow in a program. Whole program path profiling~\cite{LARUS:PLDI98} extend path profiling to analyze the program's entire control flow, including loop iterations and inter-procedural paths. We believe that unless we allow for code duplication, path profiling is too excessive of information for code layout. With code duplication, different hot paths which are executed in different contexts (and may possibly share code fragments among each other), could be laid out in different parts of the code~\cite{MuellerW:PLDI95}. However, because of the expensive compulsory misses in lower level caches and TLBs, we expect that code duplication harms the performance more than it benefits.

\section{Summary}
\label{sec:summary}
In this paper, we introduced a new technique for inter-procedural basic block layout including path-cover based BB chaining and distance-based BB collocation. Our technique achieves a performance improvement of 10\% over five widely-used programs. The improvement is primarily due to the finer grain splitting of functions and the optimal code collocation, enabled by our distance-sensitive code collocation framework.



\end{document}